\documentclass{sig-alternate}

\usepackage{amsmath}
\usepackage{amssymb}
\usepackage[np,autolanguage]{numprint}
\usepackage{units}
\usepackage{balance}

\usepackage{algorithm} 
\usepackage[noend]{algorithmic} 

\usepackage[pdftex,pdfpagelabels=false]{hyperref} 
\hypersetup{
    unicode=false,          
    pdftoolbar=true,        
    pdfmenubar=true,        
    pdffitwindow=true,      
    pdfstartview={FitV},    
    pdftitle={Costing Generated Runtime Execution Plans for Large-Scale Machine Learning Programs},    
    pdfauthor={}, 
    pdfsubject={},   
    pdfcreator={},   
    pdfproducer={}, 
    pdfkeywords={}, 
    pdfnewwindow=true,      
    bookmarksnumbered=true, 
    bookmarksopen=true,     
    bookmarksopenlevel=2,   
    colorlinks=false,        
    linkcolor=red,        
    citecolor=green,         
    filecolor=black,        
    urlcolor=black          
}

\newcommand{\card}[1]{\lvert #1\rvert}
\newcommand{\num}[1]{\numprint{#1}}

\newcommand{\s}{\unit{\,s}}
\newcommand{\mb}{\unit{\,MB}}
\newcommand{\gb}{\unit{\,GB}}
\newcommand{\tb}{\unit{\,TB}}

\newcommand{\mbs}{\,\unitfrac{\mb}{s}}

\newenvironment{itemize*}{\begin{itemize}\setlength{\itemsep}{2.5pt}\setlength{\parskip}{0pt}\setlength{\parsep}{0pt}}{\end{itemize}}
\newcommand{\eat}[1]{}

\begin{document}

\title{Costing Generated Runtime Execution Plans for\\ Large-Scale Machine Learning Programs}

\numberofauthors{1}
\author{
  \alignauthor Matthias Boehm\\~\\
  \affaddr{IBM Research -- Almaden;~~San Jose, CA, USA}\\
	\affaddr{mboehm@us.ibm.com}\\~\\
	\affaddr{\today}\\
}

\maketitle

\pagenumbering{arabic}

\sloppy

\begin{abstract}
Declarative large-scale machine learning (ML) aims at the specification of ML algorithms in a high-level language and automatic generation of hybrid runtime execution plans ranging from single node, in-memory computations to distributed computations on MapReduce (MR) or similar frameworks like Spark. The compilation of large-scale ML programs exhibits many opportunities for automatic optimization. Advanced cost-based optimization techniques require---as a fundamental precondition---an accurate cost model for evaluating the impact of optimization decisions. 
In this paper, we share insights into a simple and robust yet accurate technique for costing alternative runtime execution plans of ML programs. Our cost model relies on generating and costing runtime plans in order to automatically reflect all successive optimization phases. Costing runtime plans also captures control flow structures such as loops and branches, and a variety of cost factors like IO, latency, and computation costs. Finally, we linearize all these cost factors into a single measure of expected execution time. Within SystemML, this cost model is leveraged by several advanced optimizers like resource optimization and global data flow optimization. We share our lessons learned in order to provide foundations for the optimization of ML programs.
\end{abstract}

\section{Introduction}
\label{sec:intro}

State-of-the-art systems for large-scale ML aim at \emph{declarative} ML with high-level languages including linear algebra, statistical functions, and ML-specific constructs. This declarative approach allows users to write their custom ML algorithms once, independent of the underlying runtime framework, data or cluster characteristics. These high-level ML programs are then automatically optimized and compiled into hybrid in-memory and distributed runtime plans. The major advantages of such a high-level language are the full flexibility to specify new or customize existing ML algorithms, physical data independence of the underlying data representation (e.g., dense/sparse, formats, matrix blocking), and both efficiency and scalability via automatic cost-based optimization.\ There are many high impact optimization opportunities like static and dynamic algebraic rewrites, matrix multiplication chain optimization, decisions between single node and distributed plans, or alternative physical operators. However, any cost-based optimization technique requires an accurate cost model for evaluating alternative plans or quantifying the impact of optimization decisions.

\textbf{Cost Model Requirements:} There are several important requirements on such a cost model for optimizing large-scale ML programs which originate from potentially distributed runtime plans and ML program characteristics.
\begin{itemize*}
\item \emph{Analytical Cost Model (R1):} We need an \emph{analytical} cost model in order to cost \emph{alternative} runtime plans. The potentially large number of alternative plans prohibits a model relying on previous or sample runs.
\item \emph{Diverse Cost Factors (R2):} Large-scale ML programs exhibit several orthogonal cost factors which all can turn into bottlenecks. This includes IO, latency, and computation costs. Simple cost models like the sum of intermediate result sizes cannot capture all. 
\item \emph{Resource Awareness (R3):} The optimization of ML programs is sensitive to available memory and parallelism. Hence, our cost model needs to be aware of cluster characteristics and resource configurations.
\item \emph{Complex Control Flow (R4):} ML programs often contain deep control flow structures of loops, branches, and function calls. Our cost model needs to be able to cost arbitrary complex programs. 
\end{itemize*}

In this paper, we share a simple and robust technique of costing generated runtime plans which is the result of several lessons we have learned applying earlier cost model versions in real-world use cases of SystemML \cite{BoehmBERRSTT14, BoehmTRSTBV14, GhotingKPRSTTV11}.

\textbf{Example ML Program for Linear Regression:} As our running example, we use a simplified version of a closed-form linear regression algorithm. Its conciseness makes it feasible to present generated runtime plans, which are rarely shown in the literature. The following DML script (w/ R-like syntax) solves an ordinary least square problem $\mathbf{y}=\mathbf{X}\beta$.
\begin{verbatim}
 1:  X = read($1);
 2:  y = read($2);
 3:  intercept = $3; lambda = 0.001;
 4:  if( intercept == 1 ) {
 5:     ones = matrix(1, nrow(X), 1); 
 6:     X = append(X, ones);
 7:  }
 8:  I = matrix(1, ncol(X), 1);
 9:  A = t(X) %*% X + diag(I)*lambda;
10:  b = t(X) %*% y;
11:  beta = solve(A, b);
12:  write(beta, $4);
\end{verbatim}
In detail, we read two matrices $\mathbf{X}$ and $\mathbf{y}$ from HDFS, where we append a column of 1s to $\mathbf{X}$ if we are asked to compute the model intercept. The core computation of this ML program (lines 9-11) then constructs and solves a linear system of equations with regularization $\lambda$. The size of the intermediate results $\mathbf{A}$ and $\mathbf{b}$ is determined by the number of features. Finally, we write the model coefficients $\beta$ to HDFS.

In the rest of this paper, we discuss runtime plans generated by SystemML for different input sizes and cluster characteristics as well as the costing of these generated plans. Selected details of the entire compilation chain are described in SystemML's architecture~\cite{GhotingKPRSTTV11}, SystemML's optimizer \cite{BoehmBERRSTT14}, and SystemML's parfor optimizer for task-parallel ML programs \cite{BoehmTRSTBV14}. We leverage SystemML's text-based \texttt{EXPLAIN} tool that allows us to capture plans at different compilation levels like HOPs and runtime plans during initial compilation, as well as HOPs and runtime plans during recompilation.

\section{Generating Runtime Plans}
\label{sec:runtime}

In this section, we discuss the basics of generating runtime plans in SystemML. All examples are created on a 1+6 node cluster, i.e., one head node of 2x4 Intel E5530\,@\,2.40\,GHz-2.66\,GHz with hyper-threading enabled and $64\gb$ RAM, as well as 6 nodes of 2x6 Intel E5-2440\,@\,2.40\,GHz-2.90\,GHz with hyper-threading enabled, $96\gb$ RAM, 12x$2\tb$ disk storage, 10Gb Ethernet. We used Hadoop 2.2.0 and a static cluster configuration with $2\gb$ max/initial JVM heap size for the client and map/reduce tasks. Our HDFS capacity was $107\tb$ (11 disks per node), and we used an HDFS block size of 128\mb. Finally, our default configurations of SystemML are 12 reducers (2x number of nodes) and a memory budget ratio of 70\% of the max heap size.

\begin{table}[!t]
\centering \small
  \vspace{-0.3cm}
  \caption{\label{tab:scenarios}Overview Scenarios of Input Sizes.}
  \begin{tabular}{|c|c|c|c|}
	  \hline
    Scenario & $\mathbf{X}$ & $\mathbf{y}$ & Input Size \\ 
    \hline
    Linreg\,DS,~~XS & $10^4 \times 10^3$ & $10^4 \times 1$  & $80\mb$ \\
    Linreg\,DS,~~XL1 & $10^8 \times 10^3$ & $10^8 \times 1$  & $800\gb$ \\
    Linreg\,DS,~~XL2 & $10^8 \times 2 \cdot 10^3$ & $10^8 \times 1$  & $1.6\tb$ \\
		Linreg\,DS,~~XL3 & $2 \cdot 10^8 \times 10^3$ & $2 \cdot 10^8 \times 1$  & $1.6\tb$ \\
		Linreg\,DS,~~XL4 & $2 \cdot 10^8 \times 2 \cdot 10^3$ & $2 \cdot 10^8 \times 1$  & $3.2\tb$ \\
    \hline
  \end{tabular}
	\normalsize
	\vspace{-0.05cm}
\end{table}

\textbf{Scenarios of Different Input Sizes:} We use scenarios of different input sizes in order to show the effect on runtime plan generation. Table~\ref{tab:scenarios} gives an overview of five scenarios ranging from very small to very large use cases. In detail, this table shows the input dimensions of $\mathbf{X}$ and $\mathbf{y}$ as well as the input data size in \texttt{binary block} format. Note that we use fully dense data sets, where the number of non-zeros is equal to the number of matrix cells. In the following, we discuss generated runtime plans of selected scenarios.

\textbf{Example HOP DAG (Scenario XS):} First of all, we have a look at generated HOP DAGs for our example ML program, which allows a natural transition from script level to the level of runtime plans. We use scenario XS with input sizes of $\mathbf{X}\mbox{:~}10^4 \times 10^3$ ($80\mb$, dense, \texttt{binary block}) and $\mathbf{y}\mbox{:~}10^4 \times 1$ ($1\mb$, dense) as well as an intercept 0. Figure~\ref{fig:hops} shows the HOP \texttt{EXPLAIN} output (after HOP rewrites, computation of memory estimates, and execution type selection). There are several noteworthy modifications compared to the original script. First, after constant folding, the branch condition (lines 4-7) became constant and hence was removed accordingly. Second, multiple rewrites transformed the expression \texttt{diag(matrix(1,...))*lambda} into \texttt{diag(matrix(lambda,...))}, which prevents one unnecessary intermediate. Third, we propagated the input dimension sizes over the entire program and computed the individual operation memory estimates (input, intermediate, and output memory requirements) accordingly. Obviously, for sparse input data, this is more challenging. Fourth, according to these memory estimates and the given memory budgets (local, remote map/reduce), we selected the execution type CP (control program), i.e., pure single node, in-memory operations for all HOPs. Apart from persistent/transient read/writes, the HOP DAG contains operators for transpose (\texttt{r(t)}), matrix multiplication (\texttt{ba(+*)}), matrix construction (\texttt{dg(rand)}), vector-to-diagonal matrix (\texttt{r(diag)}), element-wise binary addition (\texttt{b(+)}), and solving a linear system of equations (\texttt{b(solve)}). This program of HOP DAGs is then compiled over LOP DAGs into a runtime program of executable program blocks and instructions. 

\begin{figure}[!t]
  \centering
\small {
\begin{verbatim}
# Memory Budget local/remote = 1434MB/1434MB/1434MB
# Degree of Parallelism (vcores) local/remote = 24/144/72
PROGRAM
--MAIN PROGRAM
----GENERIC (lines 1-3) [recompile=false]
------(10) PRead X [1e4,1e3,1e3,1e3,1e7] [76MB] CP
------(11) TWrite X (10) [1e4,1e3,1e3,1e3,1e7] [76MB] CP
------(21) PRead y [1e4,1,1e3,1e3,1e4] [0MB] CP
------(22) TWrite y (21) [1e4,1,1e3,1e3,1e4] [0MB] CP
------(24) TWrite intercept [0,0,-1,-1,-1] [0MB] CP
------(26) TWrite lambda [0,0,-1,-1,-1] [0MB] CP
----GENERIC (lines 8-12) [recompile=false]
------(42) TRead X [1e4,1e3,1e3,1e3,1e7] [76MB] CP
------(52) r(t) (42) [1e3,1e4,1e3,1e3,1e7] [153MB] CP
------(53) ba(+*) (52,42) [1e3,1e3,1e3,1e3,-1] [168MB] CP
------(50) u(ncol) (42) [0,0,-1,-1,-1] [0MB] CP
------(71) dg(rand) (50) [1e3,1,1e3,1e3,1e3] [0MB] CP
------(54) r(diag) (71) [1e3,1e3,1e3,1e3,1e3] [0MB] CP
------(57) b(+) (53,54) [1e3,1e3,1e3,1e3,-1] [15MB] CP
------(43) TRead y [1e4,1,1e3,1e3,1e4] [0MB] CP
------(59) ba(+*) (52,43) [1e3,1,1e3,1e3,-1] [76MB] CP
------(60) b(solve) (57,59) [1e3,1,1e3,1e3,-1] [15MB] CP
------(66) PWrite beta (60) [1e3,1,-1,-1,-1] [0MB] CP
\end{verbatim}
\vspace{-0.5cm}
\caption{\label{fig:hops}Example HOP DAG, Scenario XS. This program has two program blocks, w/ one HOP DAG per block. Every HOP shows its ID, operation, child IDs, output sizes (number of rows/columns, row/column block sizes, number of non-zeros), operation memory estimate, and selected execution type.}
}
\vspace{-0.1cm}
\end{figure}

\begin{figure*}[!t]
  \centering
\small {
\begin{verbatim}
PROGRAM ( size CP/MR = 34/0 )
--MAIN PROGRAM
----GENERIC (lines 1-3) [recompile=false]
------CP createvar pREADX ./mboehm/cost/X false binaryblock 10000 1000 1000 1000 10000000
------CP createvar pREADy ./mboehm/cost/y false binaryblock 10000 1 1000 1000 10000
------CP assignvar 0.SCALAR.INT.true intercept.SCALAR.INT
------CP assignvar 0.0010.SCALAR.DOUBLE.true lambda.SCALAR.DOUBLE
------CP cpvar pREADX X
------CP cpvar pREADy y
----GENERIC (lines 8-12) [recompile=false]
------CP createvar _mVar2 scratch_space//_p4140352_9.1.70.96//_t0/temp1 true binaryblock 1000 1000 1000 1000 -1
------CP tsmm X.MATRIX.DOUBLE _mVar2.MATRIX.DOUBLE LEFT
------CP createvar _mVar3 scratch_space//_p4140352_9.1.70.96//_t0/temp2 true binaryblock 1000 1 1000 1000 1000
------CP rand 1000 1 1000 1000 0.0010 0.0010 1.0 -1 uniform _mVar3.MATRIX.DOUBLE
------CP createvar _mVar4 scratch_space//_p4140352_9.1.70.96//_t0/temp3 true binaryblock 1 10000 1000 1000 10000
------CP r' y.MATRIX.DOUBLE _mVar4.MATRIX.DOUBLE
------CP createvar _mVar5 scratch_space//_p4140352_9.1.70.96//_t0/temp4 true binaryblock 1000 1000 1000 1000 1000
------CP rdiag _mVar3.MATRIX.DOUBLE _mVar5.MATRIX.DOUBLE
------CP createvar _mVar6 scratch_space//_p4140352_9.1.70.96//_t0/temp5 true binaryblock 1 1000 1000 1000 -1
------CP ba+* _mVar4.MATRIX.DOUBLE X.MATRIX.DOUBLE _mVar6.MATRIX.DOUBLE
------CP createvar _mVar7 scratch_space//_p4140352_9.1.70.96//_t0/temp6 true binaryblock 1000 1000 1000 1000 -1
------CP + _mVar2.MATRIX.DOUBLE _mVar5.MATRIX.DOUBLE _mVar7.MATRIX.DOUBLE
------CP createvar _mVar8 scratch_space//_p4140352_9.1.70.96//_t0/temp7 true binaryblock 1000 1 1000 1000 -1
------CP r' _mVar6.MATRIX.DOUBLE _mVar8.MATRIX.DOUBLE
------CP createvar _mVar9 scratch_space//_p4140352_9.1.70.96//_t0/temp8 true binaryblock 1000 1 1000 1000 -1
------CP solve _mVar7.MATRIX.DOUBLE _mVar8.MATRIX.DOUBLE _mVar9.MATRIX.DOUBLE
------CP write _mVar9.MATRIX.DOUBLE ./mboehm/cost/b.SCALAR.STRING.true textcell.SCALAR.STRING.true
\end{verbatim}
}
\vspace{-0.5cm}
\caption{\label{fig:runtime1}Example Runtime Plan, Scenario XS (same structure and characteristics as described for Figure~\ref{fig:runtime2}).}
\vspace{0.3cm}
\end{figure*}

\begin{figure*}[!t]
  \centering
\small {
\begin{verbatim}
PROGRAM ( size CP/MR = 29/1 )
--MAIN PROGRAM
----GENERIC (lines 1-3) [recompile=false]
------CP createvar pREADX ./mboehm/cost/X false binaryblock 100000000 1000 1000 1000 100000000000
------CP createvar pREADy ./mboehm/cost/y false binaryblock 100000000 1 1000 1000 100000000
------CP assignvar 0.SCALAR.INT.true intercept.SCALAR.INT
------CP assignvar 0.0010.SCALAR.DOUBLE.true lambda.SCALAR.DOUBLE
------CP cpvar pREADX X
------CP cpvar pREADy y
----GENERIC (lines 8-12) [recompile=true]
------CP createvar _mVar2 scratch_space//_p4149973_9.1.70.96//_t0/temp1 true binaryblock 1000 1 1000 1000 1000
------CP rand 1000 1 1000 1000 0.0010 0.0010 1.0 -1 uniform _mVar2.MATRIX.DOUBLE
------CP createvar _mVar3 scratch_space//_p4149973_9.1.70.96//_t0/temp2 true binaryblock 100000000 1 1000 1000 100000000
------CP partition y.MATRIX.DOUBLE _mVar3.MATRIX.DOUBLE ROW_BLOCK_WISE_N
------CP createvar _mVar4 scratch_space//_p4149973_9.1.70.96//_t0/temp3 true binaryblock 1000 1000 1000 1000 1000
------CP rdiag _mVar2.MATRIX.DOUBLE _mVar4.MATRIX.DOUBLE
------CP createvar _mVar5 scratch_space//_p4149973_9.1.70.96//_t0/temp4 true binaryblock 1000 1000 1000 1000 -1
------CP createvar _mVar6 scratch_space//_p4149973_9.1.70.96//_t0/temp5 true binaryblock 1000 1 1000 1000 -1
------MR-Job[
----------  jobtype        = GMR
----------  input labels   = [X, _mVar3]
----------  recReader inst =
----------  rand inst      =
----------  mapper inst    = MR tsmm 0.MATRIX.DOUBLE 2.MATRIX.DOUBLE LEFT, MR r' 0.MATRIX.DOUBLE 3.MATRIX.DOUBLE, 
----------                   MR mapmm 3.MATRIX.DOUBLE 1.MATRIX.DOUBLE 4.MATRIX.DOUBLE RIGHT_PART false
----------  shuffle inst   =
----------  agg inst       = MR ak+ 2.MATRIX.DOUBLE 5.MATRIX.DOUBLE true NONE, 
----------                   MR ak+ 4.MATRIX.DOUBLE 6.MATRIX.DOUBLE true NONE
----------  other inst     =
----------  output labels  = [_mVar5, _mVar6]
----------  result indices = ,5,6
----------  num reducers   = 12
----------  replication    = 1 ]
------CP createvar _mVar7 scratch_space//_p4149973_9.1.70.96//_t0/temp6 true binaryblock 1000 1000 1000 1000 -1
------CP + _mVar5.MATRIX.DOUBLE _mVar4.MATRIX.DOUBLE _mVar7.MATRIX.DOUBLE
------CP createvar _mVar8 scratch_space//_p4149973_9.1.70.96//_t0/temp7 true binaryblock 1000 1 1000 1000 -1
------CP solve _mVar7.MATRIX.DOUBLE _mVar6.MATRIX.DOUBLE _mVar8.MATRIX.DOUBLE
------CP write _mVar8.MATRIX.DOUBLE ./mboehm/cost/b.SCALAR.STRING.true textcell.SCALAR.STRING.true
\end{verbatim}
}
\vspace{-0.5cm}
\caption{\label{fig:runtime2}Example Runtime Plan, Scenario XL1 (simplified runtime plan, where we removed \texttt{rmvar} (remove variable) instructions which follow directly after the last usage of related intermediates; instructions show their execution type, operation code, input variables, output variable, and instruction-specific arguments.).}
\vspace{-0.1cm}
\end{figure*}

\textbf{Example Runtime Programs (Scenario XS):} Given the described program of HOP DAGs, we now can discuss runtime plan generation. We first look at the small scenario XS ($80\mb$) due to its simple translation. Figure~\ref{fig:runtime1} shows the generated runtime plan where we also see additional optimizer choices. First, for $\mathbf{X}^\top\mathbf{X}$ (HOP 53), we selected the physical operator \texttt{tsmm} (transpose-self matrix multiply) in order to exploit the unary input characteristic and the known result symmetry which allows to do only half the computation. Second, we applied a specific HOP-LOP rewrite, transforming $\mathbf{X}^\top\mathbf{y}$ (HOP 59) into $(\mathbf{y}^\top\mathbf{X})^\top$ in order to prevent the transpose of $\mathbf{X}$. This is done during LOP construction, because it exhibits additional memory constraints what we will discuss later in more detail. Note that we also compile size information into the runtime plan in order to provide operations with all available meta data.

\textbf{Example Runtime Program (Scenario XL1:)} We now also discuss a larger scenario XL1 ($800\gb$). For this scenario, memory estimates of HOPs 52, 53, and 59 are >$1\tb$, which is larger than the local memory budget of $\num{1434}\mb$ and hence, we select the execution type MR for these operators. Figure~\ref{fig:runtime2} shows the generated runtime plan that accordingly includes a generated MR-job instruction. There are again several interesting decisions being made here. First, we generated a hybrid runtime plan, where only operations on $\mathbf{X}$ are scheduled to MR while all other operations remain in CP. Second, we see important operator selection decisions. For $\mathbf{X}^\top\mathbf{X}$ (HOP 53), we selected again a \texttt{tsmm} MR operator but with final aggregation (\texttt{ak+}, aggregate kahan plus \cite{TianTR12}) in order to aggregate partial mapper results. This aggregation instruction is transparently used in the combiner as well. For $\mathbf{X}^\top\mathbf{y}$ (HOP 59), we selected a so-called \texttt{mapmm} (broadcast matrix multiplication), which broadcasts the smaller input through distributed cache. Similar to \texttt{tsmm}, we also have a final aggregation for this operator. Third, in contrast to Scenario XS, we did not apply the $(\mathbf{y}^\top\mathbf{X})^\top$ rewrite and hence also execute the transpose as an MR instruction. The reason for this is that the new transpose of $\mathbf{y}$ would exceed the local memory budget and hence spawn an individual MR job with related latency. Fourth, we see that our piggybacking algorithm (that packs MR operations into a minimal number of MR jobs) was able to pack all these operations into a single MR job which (1) shares the scan of $\mathbf{X}$, and prevents the materialization of $\mathbf{X}^\top$. Sixth, we decided for a CP partitioning operation of the broadcast $\mathbf{y}$ in order to reduce unnecessarily large costs for reading $\mathbf{y}$ into every task (w/o partitioning and w/o JVM reuse, we would read $800\mb$ per task input split of $128\mb$). Partitions (of $32\mb$) are read on demand but never evicted to prevent repeated partition reads. 
  
\textbf{Discussion Further Runtime Plans (Scenarios XL2/XL3/XL4):} We now discuss the even larger scenarios XL2-XL4 which all require the optimizer to generate runtime plans that exhibit very different characteristics than XL1. First, in scenario XL2, $\mathbf{X}$ has $\num{2000}$ columns which is larger than the configured block size of $\num{1000}$. This prevents the optimizer from selecting a map-side \texttt{tsmm} operator because it requires to see entire rows of the input matrix. We select an \texttt{cpmm} operator \cite{GhotingKPRSTTV11} instead, which requires two MR jobs. This implies that we have to shuffle $\mathbf{X}$ and a smaller degree of parallelism for the matrix multiplication. Piggybacking now also replicates the transpose of $\mathbf{X}$ into both jobs in order to prevent materializing the intermediate of $\mathbf{X^\top}$. Second, for scenario XL3, $\mathbf{X}$ and $\mathbf{y}$ have $2\cdot 10^8$ rows. This means that $\mathbf{y}$ is already $1.6\gb$, which is larger than the given map-task memory budget of $\num{1434}\mb$ and hence we generate a \texttt{cpmm} instead of the \texttt{mapmm}. Similar to scenario XL2, this leads to three MR jobs. Note that this decision is very sensitive to the cluster configuration (memory budget of map tasks in this case) and there are many operators that exhibit similar memory or block size constraints. Third, scenario XL4 combines the characteristics of XL2 and XL3 which leads to \texttt{cpmm} operators for both matrix multiplications but piggybacking generated again just three MR jobs because both aggregations are packed into a shared job.

To summarize, even for a very simple script, we see major plan changes for different data sizes and cluster characteristics. Optimization decisions of several compilation steps effect each other and contribute to the final runtime plan. The bottom line is that only generated runtime plans include all required information to evaluate cost factors like IO, latency, and computation costs. It is important to note, that generating runtime plans from HOP DAGs is rather efficient (<$0.5\,ms$ for common DAG sizes), which makes the generation and costing of runtime plans feasible.

\section{Costing Runtime Plans}

In this section, we now discuss how to cost generated runtime plans which automatically reflects all optimization phases. Given a runtime plan $P$ (with size information), we use a white-box cost model to compute the costs $C(P, cc)$ as estimated execution time of $P$ given the cluster configuration $cc$. This time-based model allows us to linearize IO, latency, and computation costs into the single cost measure (see R2). In contrast to related work of MR job tuning, it also gives us an \emph{analytical cost model} for entire ML programs (see R1 and R4) because it does not rely on profiling runs, and the runtime plans covers the entire control flow as well. Finally, our approach is also aware of available resources (see R3) because the compiler already respects all memory constraints when generating runtime plans, and we explicitly take the degree of parallelism into account. 

\subsection{Basic Notation}

Before we can describe the actual cost estimator skeleton, we need some basic notion. The runtime plan $P$ consists of a hierarchy of program blocks $b_i \in B$ and instructions $inst_i \in I$. A matrix $\mathbf{X}$ is described by size information of rows $m$, columns $n$, and sparsity $s$. We define $s = nnz(\mathbf{X})/(m \cdot n)$, where $nnz$ denotes the number of non-zero values. This information allows us to compute size estimates of in-memory matrices $\hat{M}(\mathbf{X})$ and serialized matrices $\hat{M}^\prime(\mathbf{X})$ (e.g., on local disk or HDFS). 
Furthermore, let $k_l$, $k_m$, and $k_r$ denote the degree of parallelism of the local control program, available map slots, and available reduce slots, respectively. In case of YARN clusters, we correct $k_m$ and $k_r$ according to the available virtual cores and memory resources of the cluster.  
Finally, let $\hat{T}(P)$ denote the estimated execution time of runtime plan $P$, which is eventually used as cost measure with $C(P, cc)=\hat{T}(P)$.

\subsection{Cost Estimator Skeleton}

The skeleton of our cost estimator recursively scans the runtime plan in execution order and tracks live variables including their sizes and in-memory state. During this single pass over the runtime program, we also compute time estimates per instruction and aggregate theses estimates accordingly to the program structure.

\textbf{Tracking Live Variable States:} Tracking sizes and in-memory state of variables is a fundamental precondition for costing individual instructions. We start with an empty symbol table. While costing the runtime plan, we maintain live variable statistics in this table. First, for each \texttt{createvar} (creates meta data handle for a matrix variable), \texttt{cpvar} (binds a variable to a variable name), \texttt{rmvar} (removes a variable), and data generating instructions like \texttt{rand} or \texttt{seq}, we accordingly modify our live variable statistics (e.g., size information). Second, we also maintain in-memory state of variables. Persistent read inputs and MR job outputs are known to be on HDFS, while all in-memory instructions change the state of their inputs and output to in-memory. This state maintenance allows us to correctly reflect required IO costs. For example, if a persistent dataset is used by two in-memory instructions, only the first instruction will pay the costs of reading the input. This approach also allows us to reason about hybrid runtime plans of CP/MR instructions, where intermediates are exchanged via HDFS.

\textbf{Time Estimate Aggregation over Control Flow:} Finally, we aggregate time estimates as we recursively iterate over the program structure. Similar to statistics aggregation in the parfor optimizer for task-parallel ML programs \cite{BoehmTRSTBV14}, we aggregate the time estimate of an program block $b$ by the sum of its childs $c(b)$ (predicates, included program blocks, instructions) due to their sequential execution with:  
\begin{equation}
  \hat{T}(b)= w_b
  \sum_{\forall c_i \in c(b)} \mbox{\hspace{-0.15cm}} \hat{T}(c_i)
  \mbox{,~~} w_b=
  \begin{cases}
    \lceil\hat{N}/k\rceil & \texttt{parfor}\\
    \hat{N}&\texttt{for,while}\\
    1/\card{c(n)} &\texttt{if}\\
    1 & \mbox{otherwise}.\\
  \end{cases}
\end{equation}
For conditional branches, the aggregate is a weighted sum of time estimates for the individual branches. For loops, we scale the time aggregate by the number of iterations; if the number of iterations is unknown (e.g., for \texttt{while} loops) we use a constant $\hat{N}$ which at least reflects that the body is executed multiple times. Note that we use additional corrections in order to account for overestimated read costs in loops, where only the first iteration reads persistent inputs. Furthermore, we also maintain function call stacks in order to prevent cycles when costing recursive functions.

This cost estimator skeleton allows the costing of arbitrary complex runtime plans including control flow structures. The actual time estimation problem then boils down to estimating execution time of a single instruction given the size and in-memory state of its input and output variables.

\subsection{Time Estimates of Instructions}

In general, we compute the time estimate of an instruction as the sum of latency, IO, and computation time based on its input and output statistics. Earlier versions of this cost model \cite{BoehmTRSTBV14} relied on profiled and trained cost functions for individual instructions. In contrast, we now use a white-box cost model based on IO bandwidth multipliers and operation-specific floating point operations in order to remove the need for cluster-specific profiling runs.

\textbf{Costing CP Instructions:} The time estimate of a CP instruction consists of IO and compute time. We estimate IO time based on variable state, size, format, and default format-specific IO bandwidths. If the state of an input is in-memory, then there is no IO time; otherwise, we compute the IO time via the serialized, format-specific size $\hat{M}^\prime(\mathbf{X})$ of this input. For example, given a $10^4 \times 10^3$ dense matrix in \texttt{binary block} format, we get $\hat{M}^\prime(\mathbf{X})=80\mb$; by weighting this with the single-threaded read bandwidth for \texttt{binary block} ($150\mbs$), we get an IO time of $0.53\s$. Compute time is estimated as the maximum of main memory IO (computed via main memory bandwidth multipliers) and instruction-specific models of required floating point operations. For example, let us use the \texttt{tsmm} (transpose-self matrix multiplication) instruction for $\mathbf{X}^\top \mathbf{X}$ that we introduced earlier. Its floating point requirements are estimated as follows:
\begin{equation}
  \mbox{FLOP}(\texttt{tsmm}_{\mbox{left}})= 
  \begin{cases}
    \mbox{MMD\_corr} \cdot m \cdot n^2 \cdot s & \texttt{dense}\\
    \mbox{MMS\_corr} \cdot m \cdot n^2 \cdot s^2 & \texttt{sparse}\\
  \end{cases}
\end{equation}
Finally, we convert the required flops into expected execution time assuming 1FLOP per cycle. For example, for $\mathbf{X}:10^4 \times 10^3$, $\mbox{MMD\_corr}=0.5$ (operation-specific correction), and a 2\,GHz processor, we get $\hat{T}(inst)=0.5\cdot10^{10} / (2\cdot10^9) = 2.5\s$. Note that our cost model consists of dozens of these white-box cost functions for all existing instructions.  

\textbf{Costing MR-Job Instructions:} The time estimate of an MR-job instruction is more complex. It consists of job and task latency, write times for in-memory variable export, map task read, compute, and write times, shuffle time, as well as reduce task read, compute, and write times. The individual IO times and computation times are estimated similar to CP instructions, but weighted with the degree of parallelism of map/reduce tasks. Note that costing needs to take the structure of the MR job into account. For example, consider a map-only job with a single \texttt{mapmm} instruction without final aggregation for $\mathbf{X}~\mathbf{v}$. This job will incur job and task latency as well as map read costs for $\mathbf{X}$ and $\mathbf{v}$, the matrix-vector computation costs, and finally the map result write costs. The sum of these map-side costs are divided by the effective degree of parallelism, which is computed via a scaled minimum of $k_m$ (available parallelism) and number of tasks ($\hat{M}^\prime(\mathbf{X})$ divided by the HDFS block size). On YARN clusters, we also take the CP/MR memory resources into account when computing the degree of parallelism. 


\subsection{Examples Runtime Plan Costing} 

Putting it all together, we now revisit the example runtime plans from Section~\ref{sec:runtime} and discuss their costing in detail.

\begin{figure}[!b]
  \centering
\small {
\begin{verbatim}
PROGRAM                             # total cost C=3.31s 
--MAIN PROGRAM                       # C=3.31s
----GENERIC (lines 1-3)               # C=2.8E-8s  
------CP createvar pREADX binaryblock  # C=[0s,  4.7E-9s]
------CP createvar pREADy binaryblock  # C=[0s,  4.7E-9s]
------CP assignvar intercept           # C=[0s,  4.7E-9s]
------CP assignvar lambda              # C=[0s,  4.7E-9s]
------CP cpvar pREADX X                # C=[0s,  4.7E-9s]
------CP cpvar pREADy y                # C=[0s,  4.7E-9s]
----GENERIC (lines 8-12)              # C=3.31s  
------CP createvar _mVar2              # C=[0s,  4.7E-9s] 
------CP tsmm X _mVar2 LEFT            # C=[0.51s, 2.32s]
------CP createvar _mVar3              # C=[0s,  4.7E-9s]
------CP rand 1000 1 _mVar3            # C=[0s,  3.7E-6s]
------CP createvar _mVar4              # C=[0s,  4.7E-9s]
------CP r' y _mVar4                   # C=[5E-4s, 5E-6s]
------CP createvar _mVar5              # C=[0s,  4.7E-9s]
------CP rdiag _mVar3 _mVar5           # C=[0s,  4.7E-7s]
------CP createvar _mVar6              # C=[0s,  4.7E-9s]
------CP ba+* _mVar4 X _mVar6          # C=[0s, 0.00465s]
------CP createvar _mVar7              # C=[0s,  4.7E-9s]
------CP + _mVar2 _mVar5 _mVar7        # C=[0s,  4.7E-4s]
------CP createvar _mVar8              # C=[0s,  4.7E-9s]
------CP r' _mVar6 _mVar8              # C=[0s,  4.7E-7s]
------CP createvar _mVar9              # C=[0s,  4.7E-9s]
------CP solve _mVar7 _mVar8 _mVar9    # C=[0s,   0.466s]
------CP write _mVar9 textcell         # C=[1E-6s, 2E-4s]
\end{verbatim}
}
\vspace{-0.5cm}
\caption{\label{fig:cost1}Simplified Plan Scenario XS w/ Costs.}
\vspace{-0.05cm}
\end{figure}

\textbf{Example Plan Costing (Scenario XS):} Figure~\ref{fig:cost1} shows a simplified runtime plan for scenario XS ($80\mb$) with annotated costs. Due to the simple program structure, the total plan execution time of $3.31\s$ is computed as the plain sum of all instruction costs (which we show as a breakdown of IO and compute time). There are a couple of interesting observations to make. First, the instruction that uses a persistent input first, pays the related IO costs (e.g., \texttt{tsmm} and \texttt{r'}), while subsequent operations on the same data (e.g., \texttt{ba+*}) do only account for compute time. Second, we see that the computation time for \texttt{tsmm} dominates the total execution time. The following heavy hitters are the initial read of $\mathbf{X}$ as well as the computation costs of \texttt{solve}.

\begin{figure}[!t]
  \centering
\vspace{0.15cm}	
\small {
\begin{verbatim}
PROGRAM                            # total cost C=606.9s 
--MAIN PROGRAM                      # C=606.9s
----GENERIC (lines 1-3)              # C=2.8E-8s  
------CP createvar pREADX binaryblock # C=[0s,  4.7E-9s]
------CP createvar pREADy binaryblock # C=[0s,  4.7E-9s]
------CP assignvar intercept          # C=[0s,  4.7E-9s]
------CP assignvar lambda             # C=[0s,  4.7E-9s]
------CP cpvar pREADX X               # C=[0s,  4.7E-9s]
------CP cpvar pREADy y               # C=[0s,  4.7E-9s]
----GENERIC (lines 8-12)             # C=606.9s
------CP createvar _mVar2 binaryblock # C=[0s,  4.7E-9s]
------CP rand 1000 1 _mVar2           # C=[0s,  3.7E-6s]
------CP createvar _mVar3             # C=[0s,  4.7E-9s]
------CP partition y _mVar3           # C=[10.2s,  6.4s]
------CP createvar _mVar4             # C=[0s,  4.7E-9s]
------CP rdiag _mVar2 _mVar4          # C=[0s,  4.7E-7s]
------CP createvar _mVar5             # C=[0s,  4.7E-9s]
------CP createvar _mVar6             # C=[0s,  4.7E-9s]
------MR-Job[ # nmap=5967 nred=1      # C=[589.8s]
----------jobtype = GMR                # latency=[144.5s]
----------inputs  = [X, _mVar3]        # hdfsread=[70.7s]
----------map     = MR tsmm 0 2,       # mapexec=[324.7s]
----------          MR r' 0 3, 
----------          MR mapmm 3 1 4     # dcread=  [12.6s]
----------shuffle =                    # shuffle= [19.7s] 
----------agg     = MR ak+ 2 5,        # redexec= [11.1s] 
----------          MR ak+ 4 6
----------outputs = [_mVar5, _mVar6]   # hdfswrite=[0.1s]
----------ret ix  = ,5,6
----------repl    = 1 ]
------CP createvar _mVar7             # C=[0s,  4.7E-9s]
------CP + _mVar5 _mVar4 _mVar7       # C=[0.05s, 5E-4s]
------CP createvar _mVar8             # C=[0s,  4.7E-9s]
------CP solve _mVar7 _mVar6 _mVar8   # C=[5E-5s,0.466s]
------CP write _mVar8 textcell        # C=[1E-6s, 2E-4s]
\end{verbatim}
}
\vspace{-0.55cm}
\caption{\label{fig:cost2}Simplified Plan Scenario XL w/ Costs.}
\vspace{-0.3cm}
\end{figure}

\textbf{Example Plan Costing (Scenario XL1):} As stated before, costing plans that include MR-job instructions is more challenging than pure CP runtime plans. Figure~\ref{fig:cost2} shows the simplified runtime plan of scenario XL1 ($800\gb$) with annotated costs. In comparison to scenario XS, there are many additional cost factors. First, cost estimates of CP instructions automatically adapt to the increased data sizes and additional operators. For example, now the \texttt{partition} instruction pays the $10.2\s$ costs for the initial read of $\mathbf{y}$. Second, the total execution time of $606.9\s$ is dominated by the costs of $589.8\s$ for the generated MR job. Several cost factors contribute to this estimate. The total estimated latency includes $20\s$ job latency plus $1.5\s$ task latency for each map/reduce tasks, normalized by the effective map and reduce degree of parallelism. Furthermore, the HFDS read costs reflect reading all map inputs, again normalized by the effective degree of parallelism. The major cost factor, however, of this compute-intensive job is the map compute time which is dominated by \texttt{tsmm}. Additional cost factors include read from distributed cache for the partitioned broadcast in \texttt{mapmm}, shuffle IO time, reduce compute time (for the final aggregations), and the final HDFS write of $\mathbf{A}$ and $\mathbf{b}$. Here, the shuffle time captures map write, actual shuffle, and reduce write/read. Third, despite the same remaining instructions as in scenario XS, we see slightly different costs (e.g., for \texttt{+} and \texttt{solve}) because by tracking the in-memory state of variables, we automatically take hybrid runtime plans with data exchange over HDFS into account as well. 

Regarding cost model accuracy, in both examples, the estimated costs were within 2x of the actual execution time. Due to simplifying assumptions and fundamental limitations, this is not given in general. However, this cost model allows for reasonable cost comparisons of complex ML programs without the need of profiling or sample runs.

\subsection{Limitations}

The presented cost model works very well in practice. However, there are also fundamental limitations.

\textbf{Unknown Size Information:} Despite techniques for propagating size information of dimensions and sparsity \cite{BoehmBERRSTT14}, there do exist cases where we are not able to determine sizes of intermediates during initial compilation. In this case, the compiler falls back to conservative but scalable plans in order to ensure plan validity. However, apart from MR job latency, we cannot fully infer IO and computation costs of affected operators in those cases which potentially leads to large underestimation. This issue is commonly addressed by making the optimizer, using the cost model, aware of unknowns, which can often even be used for pruning.

\textbf{Buffer Pool Behavior:} Our cost model only partially considers buffer pool evictions which may contribute to the overall program costs. In order to fully address this, we would need a white-box model of the buffer pool eviction algorithm and extend the tracking of live variables. For the sake of simplicity, we currently view the buffer pool as black box and only consider its total size. In practice, this is acceptable since buffer pool evictions usually account for a small fraction of the total execution time. 

\textbf{Unknown Conditional Control Flow:} Many ML programs contain conditional control flow in terms of loops with unknown number of iterations, branches, and recursive function calls. Especially for convergence-based ML algorithms, the number of iterations until convergence is generally unknown. Our heuristic of predefined constants clearly can fail there but at least reflects that the loop body is executed repeatedly. This already allows for optimization techniques like code motion or caching decisions. There is also existing work on estimating the number of iterations until convergence, which is an interesting direction for future extensions.

\section{Conclusions}

To summarize, our simple and robust cost model allows the costing of generated runtime plans for ML programs. This model automatically reflects all optimization decisions of the entire compilation chain. Most importantly it provides an analytical cost model for alternative plans without the need for profiling or sample runs. It also captures all relevant cost factors, is aware of data and cluster characteristics, and can be used for arbitrary complex ML programs. 

\bibliographystyle{abbrv}
\bibliography{tr_costmodel2015}

\enlargethispage{\baselineskip}

\end{document}